\documentclass[aip,amsmath,amssymb,reprint]{revtex4-1}
\title{This is \texttt{main.tex}}
\draft
\usepackage{graphicx}  
\usepackage{dcolumn}
\usepackage{amsmath}
\usepackage{makecell}
\usepackage{bm}        
\usepackage{amssymb}   
 \usepackage{placeins}
\usepackage[utf8]{inputenc}
\usepackage[T1]{fontenc}
\usepackage{mathptmx}
\usepackage{dblfloatfix}
\usepackage{float}
\usepackage{color}
\usepackage{xr}
\usepackage[most]{tcolorbox}
\usepackage{lipsum}   


\makeatletter

\newcommand*{\addFileDependency}[1]{
\typeout{(#1)}
\@addtofilelist{#1}
\IfFileExists{#1}{}{\typeout{No file #1.}}
}\makeatother

\newcommand*{\myexternaldocument}[1]{%
\externaldocument{#1}%
\addFileDependency{#1.tex}%
\addFileDependency{#1.aux}%
}

\myexternaldocument{SI}

\begin{document}

\title{Auxetic behaviour in crystals of hard polyhedra}

\author{Rinske M. Alkemade}
\affiliation{Soft Condensed Matter and Biophysics, Debye Institute for Nanomaterials Science, Utrecht University, Utrecht, Netherlands }
\author{Silvana A. Caipa Cure}
\affiliation{Soft Matter Physics, Huygens-Kamerlingh Onnes Laboratory, Leiden University, The Netherlands}
\author{Alptu\u{g} Ulug\"ol}
\affiliation{Soft Condensed Matter and Biophysics, Debye Institute for Nanomaterials Science, Utrecht University, Utrecht, Netherlands }
\affiliation{Institute of Physics, University of Augsburg, Universit\"{a}tsstr. 1, 86159 Augsburg, Germany}
\author{Andrea Plati}
\affiliation{
Universit\'e Paris-Saclay, CNRS, Laboratoire de Physique des Solides, 91405 Orsay, France
}
\author{Vera Belde}
\affiliation{Soft Condensed Matter and Biophysics, Debye Institute for Nanomaterials Science, Utrecht University, Utrecht, Netherlands }
\author{Haadi Naqvi}
\affiliation{Soft Condensed Matter and Biophysics, Debye Institute for Nanomaterials Science, Utrecht University, Utrecht, Netherlands }
\author{Felix Verbiest}
\affiliation{Soft Condensed Matter and Biophysics, Debye Institute for Nanomaterials Science, Utrecht University, Utrecht, Netherlands }
\author{Giuseppe Foffi}
\affiliation{
Universit\'e Paris-Saclay, CNRS, Laboratoire de Physique des Solides, 91405 Orsay, France
}
\author{Daniela Kraft}
\affiliation{Soft Matter Physics, Huygens-Kamerlingh Onnes Laboratory, Leiden University, The Netherlands}
\author{Frank Smallenburg}
\affiliation{
Universit\'e Paris-Saclay, CNRS, Laboratoire de Physique des Solides, 91405 Orsay, France
}
\author{Laura Filion}
\affiliation{Soft Condensed Matter and Biophysics, Debye Institute for Nanomaterials Science, Utrecht University, Utrecht, Netherlands }
\begin{abstract}
Auxetic materials -- systems that, when subjected to a compression in one direction, also compress in one or more perpendicular directions -- have intrigued researchers for decades due to their counterintuitive mechanical properties. Their unique behaviour gives auxetic materials potential for a wide range of applications such as shock absorbers, and electrodes in piezoelectric sensors.  Most known auxetic materials are realized by connecting rigid, anisotropic units in a hinging manner, which are systems that often are hard to realize experimentally on the microscopic scale. Here, we explore the elastic behaviour of six crystals composed of discrete space-filling hard polygons or polyhedra. We show that some of these systems show partial auxetic behaviour, emerging from the interplay of entropy and geometry alone. To demonstrate the feasibility and robustness of this phenomenon, we create two experimental realizations of square-shaped particles, spanning both colloidal particles driven by Brownian motion and granular particles driven by external vibrations, and confirm the emergence of auxeticity in both cases.
\end{abstract}

\maketitle

\section{Introduction}

The overarching goal in the field of metamaterials is the targeted design of new materials with specific macroscopic properties \cite{zheludev2012metamaterials}. Such properties can include negative-index of refraction\cite{smith2000composite,shelby2001experimental}, negative Hall coefficient\cite{briane2009homogenization}, and unusual responses to mechanical deformation\cite{lakes1987foam, lakes2001extreme, wang2023non}. This latter category also encompasses negative Poisson ratios: the property that a material, when subjected to a compression in one direction, also compresses in one or more perpendicular directions. This runs counter to the behavior of most materials, which ``bulge out'' when squeezed, and gives so called auxetic materials potential for a wide range of applications\cite{evans2000auxetic, yang2004review} such as shock absorbers\cite{wang2016negative}, acoustic wave suppression\cite{li2025application} and electrodes in piezoelectric sensors\cite{hadjigeorgiou2004use}.
\begin{figure}[t]
    \centering
    \includegraphics[width=0.85\linewidth]{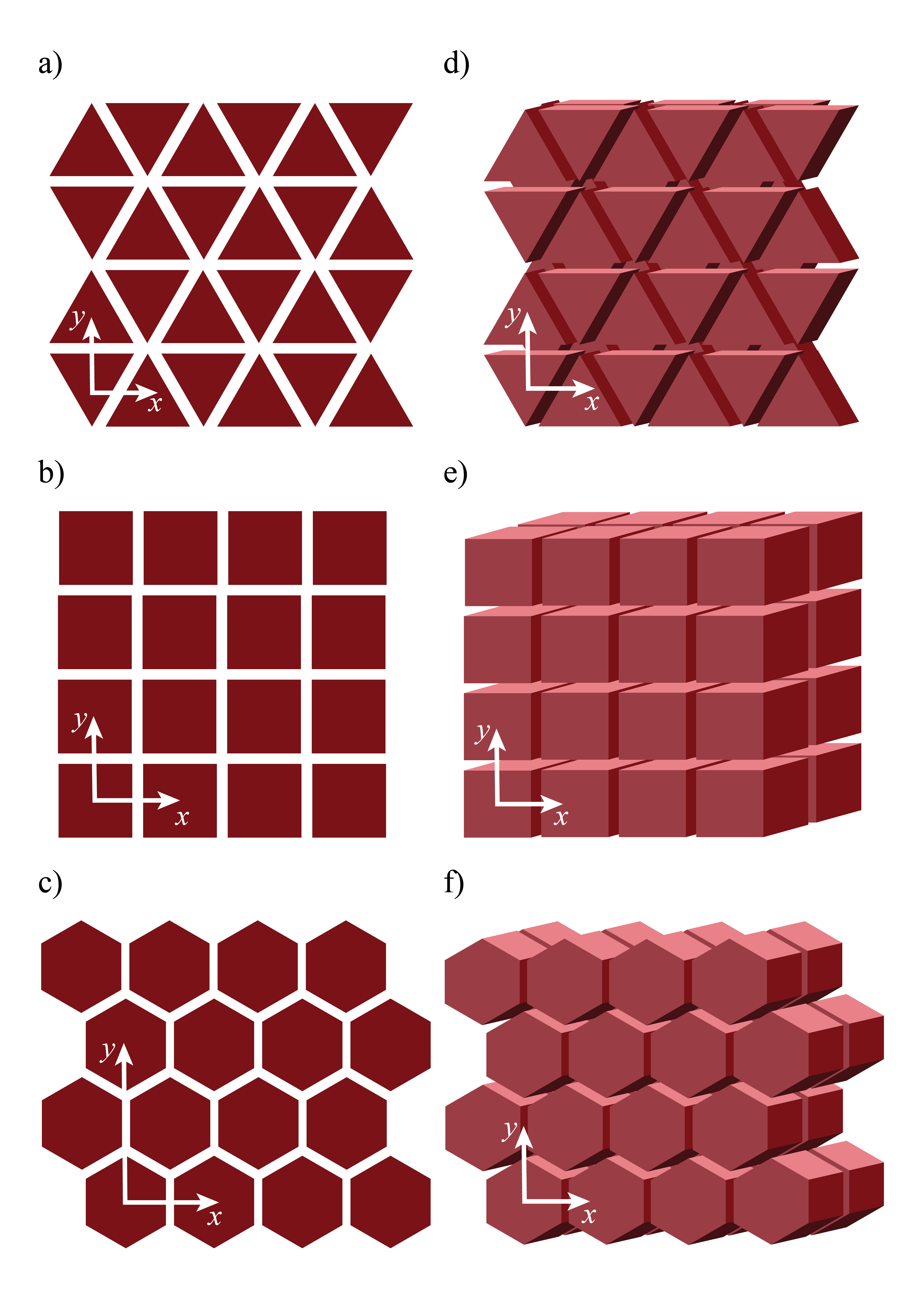}
    \caption{The six crystals under consideration in this paper, consisting of a) hard triangles, b) hard squares and c) hard hexagons in two dimensions and d) hard triangular prisms, e) hard cubes and f) hexagonal prisms in three dimensions.}
    \label{fig:difcrystals}
\end{figure}

One of the earliest mechanical realizations of auxetic behaviour -- although that term was only introduced later -- can be found in the pioneering work of Resch in the 1960s, in which he showed that hingedly connected rigid polygons  exhibit a negative Poisson ratio\cite{resch1965geometrical}. However, the field really started to advance after the seminal work of Lakes\cite{lakes1987foam} who introduced auxetic foams. Since then, a zoo of auxetic materials has been developed and constructed. Most known auxetic materials are realized by connecting rigid, anisotropic units in a hinging manner, allowing for the leverage of a single soft mode that connects the structure in two perpendicular dimensions\cite{andrade2018extreme, grima2019novel,wang2024poisson}. Auxetic materials that fall in this category are for example constructed by using i) re-entrant building blocks \cite{lakes1987foam, friis1988negative,chen1989dynamic}  ii) chiral building blocks \cite{prall1997properties} and iii) rotating building blocks\cite{grima2000auxetic, melioPivotingColloidalAssemblies2026}. 
 
Although intriguing, realizing these network-based architectures at the microscale can be challenging. A possible alternative can be found in auxetic materials consisting of discrete objects, as these systems are often easier to realize at small length scales. For example, most crystals with a cubic-shaped lattice cell have been shown to possess a negative Poisson ratio along non-axial directions \cite{baughman2000negative}, a characteristic that can be explained by the geometry and the elastic anisotropy of the system. Systems of non-convex hard particles\cite{wojciechowski1989negative,wojciechowski2003elastic,tretiakov2007poisson,tretiakov2009negative}, which are governed purely by geometry and entropy, also exhibit negative Poisson ratios due to geometrical locking of the individual shapes. 

These results raise the question whether the interplay between geometry and entropy can lead to negative Poisson ratios in even simpler systems. To this end, we explore the elastic behaviour of six crystals composed of space-filling hard polyhedra (see Fig.~\ref{fig:difcrystals}), which are among the geometrically simplest systems imaginable. We find that several of these crystals exhibit partial auxetic behaviour. Moreover, we confirm this prediction experimentally using two experimental setups spanning all the way from the colloidal to the granular scale -- demonstrating that entropy-driven auxeticity is a robust phenomenon in hard polyhedra.

\begin{figure}
        \begin{minipage}[t]{0.99\linewidth}
        \vspace{0pt}
        \raggedright a)\\
        \includegraphics[width=\linewidth]{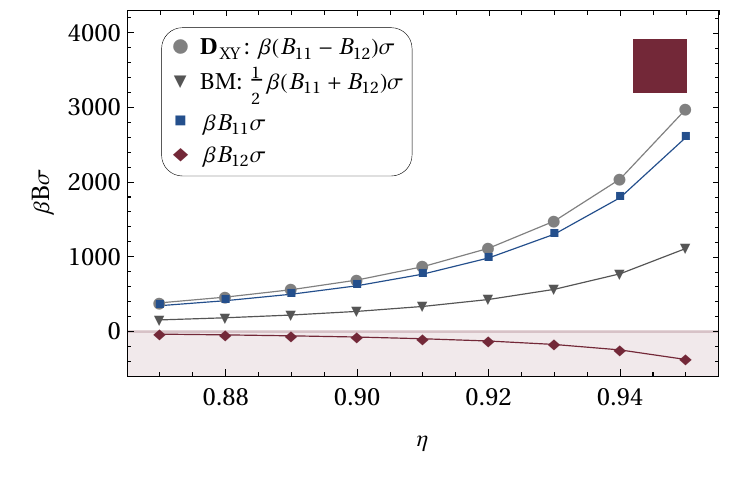}
        \end{minipage}
        \\
        \begin{minipage}[t]{0.99\linewidth}
        \vspace{0pt}
        \raggedright b)\\
        \includegraphics[width=\linewidth]{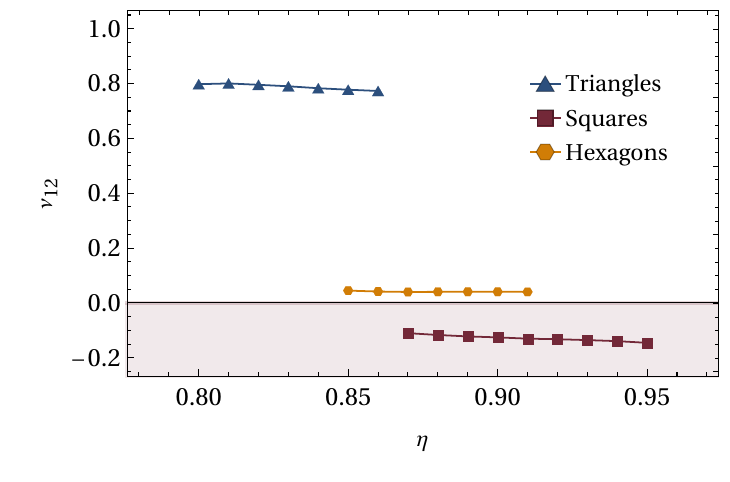}
        \end{minipage}
        
    \caption[width=1\linewidth]{a) Linear contributions of the stress-strain curves for a system subjected to a $\mathbf{D}_{XY}$-deformation (see SI), together with the Bulk modulus (BM) of a square crystal, both plotted in gray, as a function of $\eta$. From this data, the values of $B_{11}$ and $B_{12}$ are extracted and displayed. b) Poisson ratio $\nu_{12}$ as a function of $\eta$ for the three two-dimensional systems considered in this paper.}
    \label{fig:2Dresultstog}
\end{figure}

\section{Elastic behaviour of crystals of hard squares}

We begin our investigation by exploring the elastic behaviour of crystals of hard squares. Specifically, we are interested in the effective elastic constants, $B_{ijkl}$, as they can be used to determine the Poisson ratio. 
These constants quantify the linear response to stress of a material under hydrostatic pressure \cite{frenkel2023understanding}, i.e.
\begin{align}
\frac{\partial \sigma_{ij}}{\partial \epsilon_{kl}}\bigg\vert_{\epsilon = 0} = C_{ijkl}-\left(\delta_{ik}\delta_{jl}+ \delta_{il}\delta_{jk} -\delta_{ij}\delta_{kl}\right)P \equiv B_{ijkl},
\label{eq:elasticconstants}
\end{align}
with $C_{ijkl}$ the elastic constants tensor, $P$ the hydrostatic pressure and $\sigma_{ij}$  and $\epsilon_{ij}$ the stress and strain in the $ij$ direction, respectively. A derivation of Eq.~\eqref{eq:elasticconstants} is provided in the Supplementary Information (SI). 
The Poisson ratio can then be expressed in terms of these effective elastic constants $\mathbf{B}$ via\cite{Timoshenko_1951aa}
\begin{align}
    \nu_{ij}  \equiv - \frac{\partial\epsilon_{ii}}{\partial\epsilon_{jj}} = - \frac{S_{iijj}\partial\sigma_{jj}}{S_{jjjj}\partial\sigma_{jj}}= - \frac{S_{iijj}}{S_{jjjj}},
    \label{eq:poisonratio3}
\end{align}
where $\mathbf{S} \equiv \mathbf{B}^{-1}$. Here, we express and extract the components of $B_{ijkl}$ in the coordinate system indicated in Fig.~\ref{fig:difcrystals}b), i.e $ijkl\in \{x,y\}$. However, since $\mathbf{B}$ fully characterizes the elastic response of the system, knowing its value for one set of principal axes is sufficient to determine $\nu_{ij}$ for any arbitrary orientation through tensor rotation. 

As explained in detail in the SI, we can extract the effective elastic constants from simulations by measuring the bulk modulus as well as applying a series of small deformations and measuring the associated stress. Here, we use event-driven molecular dynamics simulations to model the systems \cite{hernandez2007discontinuous, smallenburg2012vacancy}.  Since we are considering a crystal with square symmetry, there are only three independent elastic constants: $B_{11}, B_{12}$ and $B_{66}$. Here, we use Voigt notation to express the effective elastic tensor. Voigt notation performs a symmetry-based contraction of indices in a symmetric tensor, thereby reducing the order of the tensor\cite{voigt1910lehrbuch} (see SI). We show the dependence of $B_{11}$ and $B_{12}$ on the packing fraction $\eta$ in Fig.~\ref{fig:2Dresultstog}a) for the crystal orientation as depicted in Fig.~\ref{fig:difcrystals}b) (the results for $B_{66}$ can be found in the SI).
As expected, the figure shows that the magnitude of the elastic constants increases with packing fraction. 

From $B_{11}$ and $B_{12}$ we can determine the Poisson ratio associated with compressing the crystal along the $x$-axis, which is given by $\nu = \frac{B_{12}}{B_{11}}$. The result is shown in Fig.~\ref{fig:2Dresultstog}b).  Interestingly, for all packing fractions studied, we find a negative Poisson ratio along this direction for the square system.

\begin{figure*}[p]
\begin{tcolorbox}[enhanced,
  colframe=black, colback=white,
  boxrule=0.4mm, arc=4mm,
  width=\linewidth, boxsep=2mm]

\begin{minipage}[c]{.5cm}
\vfill
\rotatebox{90}{\centering I. SIMULATIONS}
\vfill
\end{minipage}%
\hfill
\begin{minipage}[c]{0.96\linewidth}
\begin{minipage}[t]{0.25\linewidth}
\raggedright a)\\
\includegraphics[height=4.3cm]{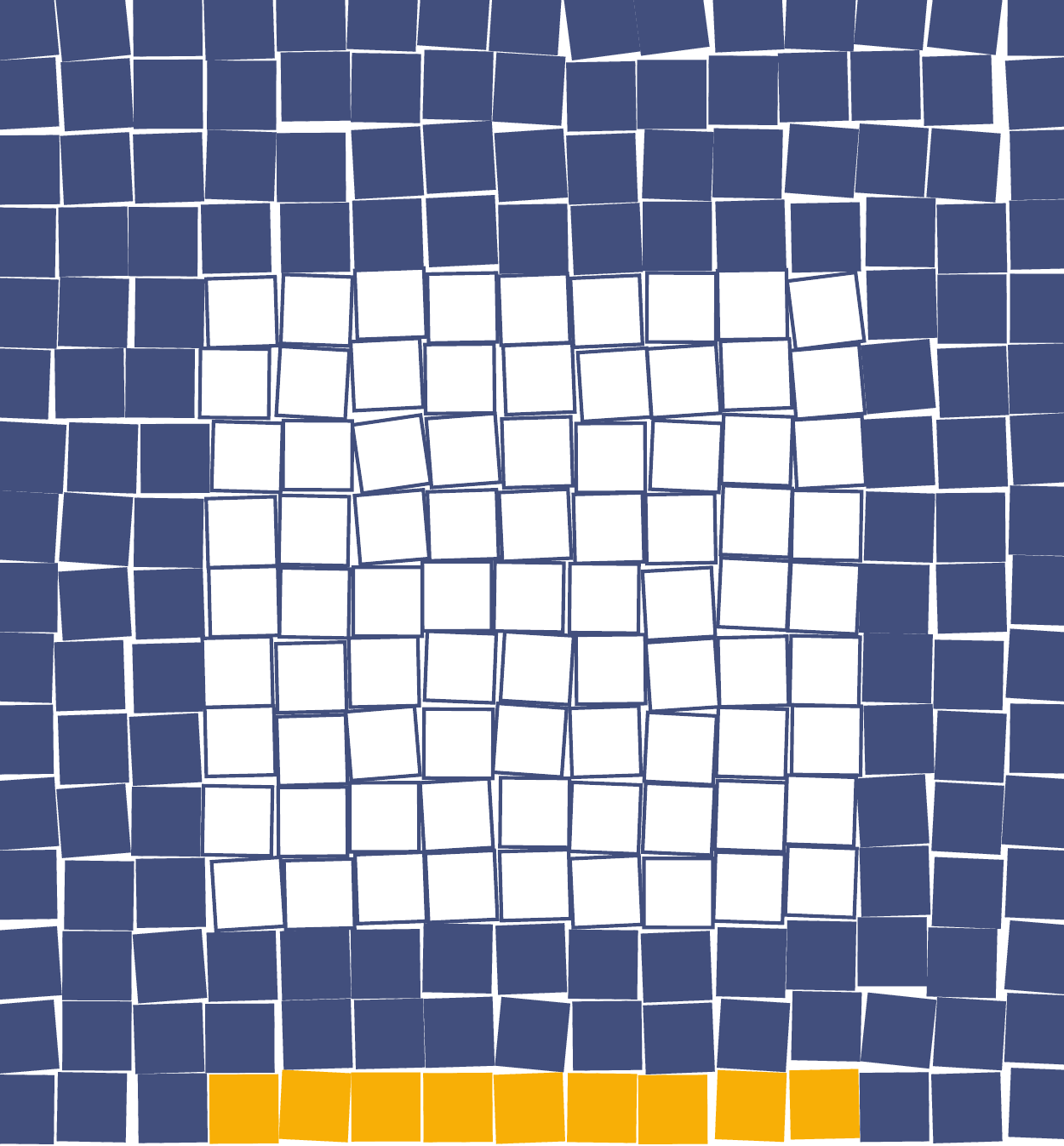}
\end{minipage}\hfill
\begin{minipage}[t]{0.25\linewidth}
\raggedright {\color{white}.}\\
\includegraphics[height=4.3
cm]{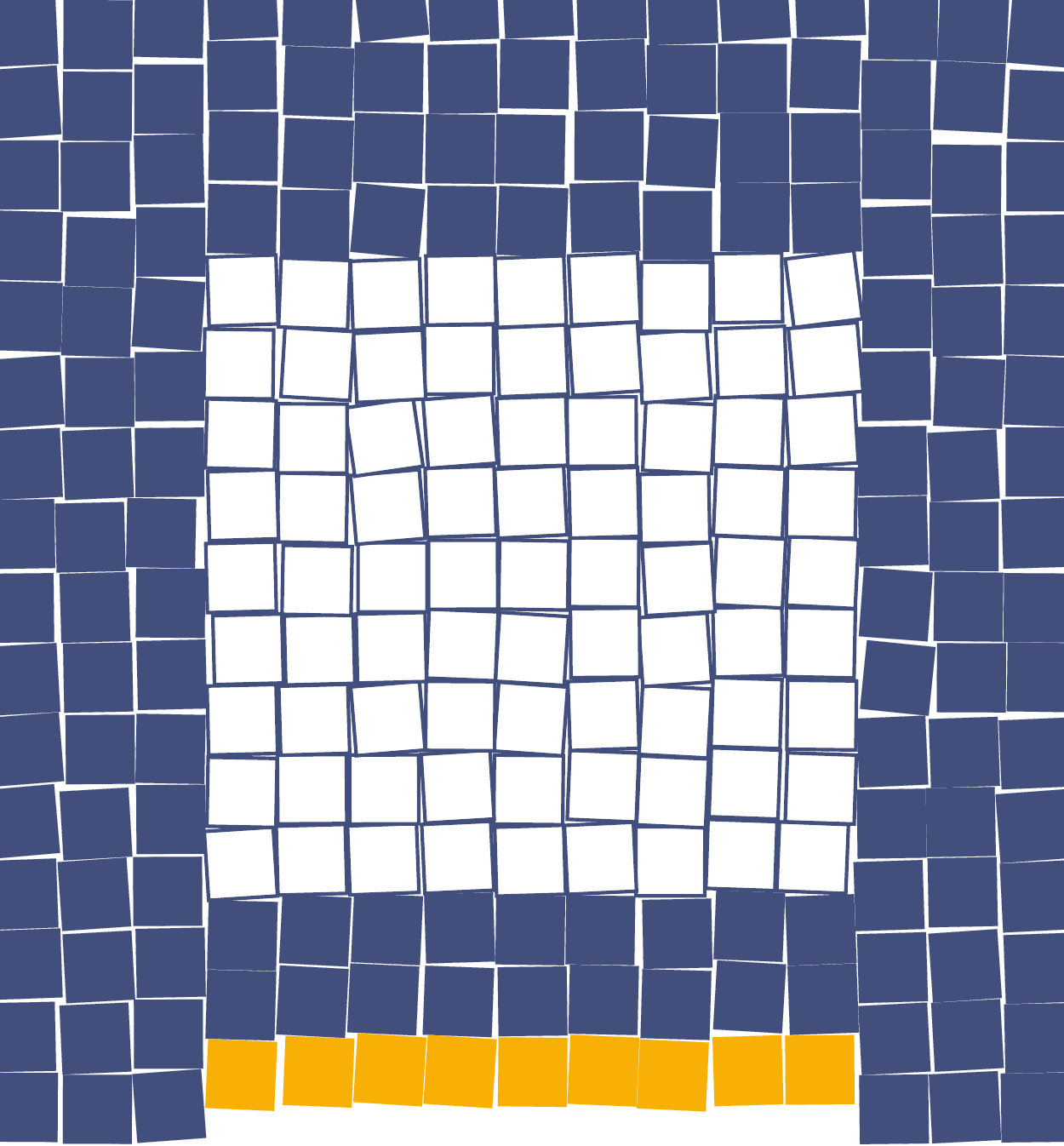}
\end{minipage}\hfill
\begin{minipage}[t]{0.45\linewidth}
\raggedright b)\\
\includegraphics[height=4.5cm]{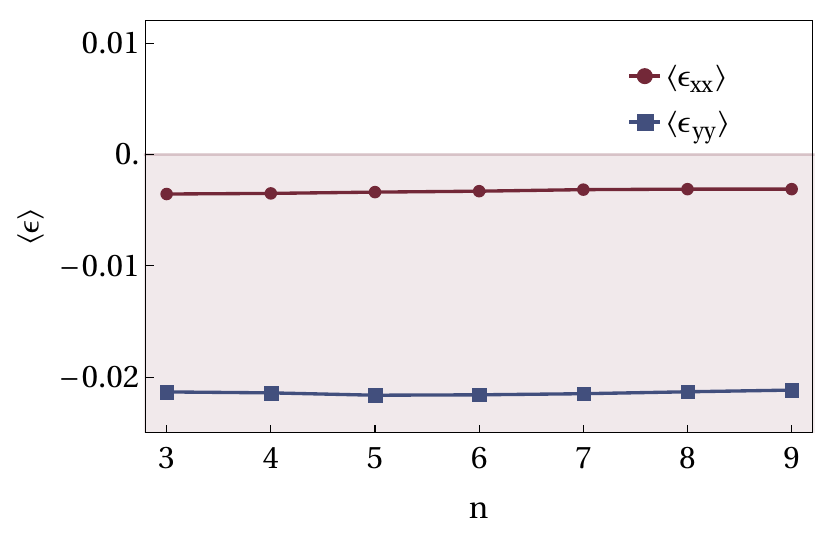}
\end{minipage}
\end{minipage}

\end{tcolorbox}

\vspace{0.3cm} 
\begin{tcolorbox}[enhanced,
  colframe=black, colback=white,
  boxrule=0.4mm, arc=4mm,
  width=\linewidth, boxsep=2mm]

\begin{minipage}[c]{.5cm}
\vfill
\rotatebox{90}{\centering II. COLLOIDAL EXPERIMENTS}
\vfill
\end{minipage}%
\hfill
\begin{minipage}[c]{0.96\linewidth}
\begin{minipage}[t]{0.30\linewidth}
\raggedright c)\\
\includegraphics[height=4.4cm]{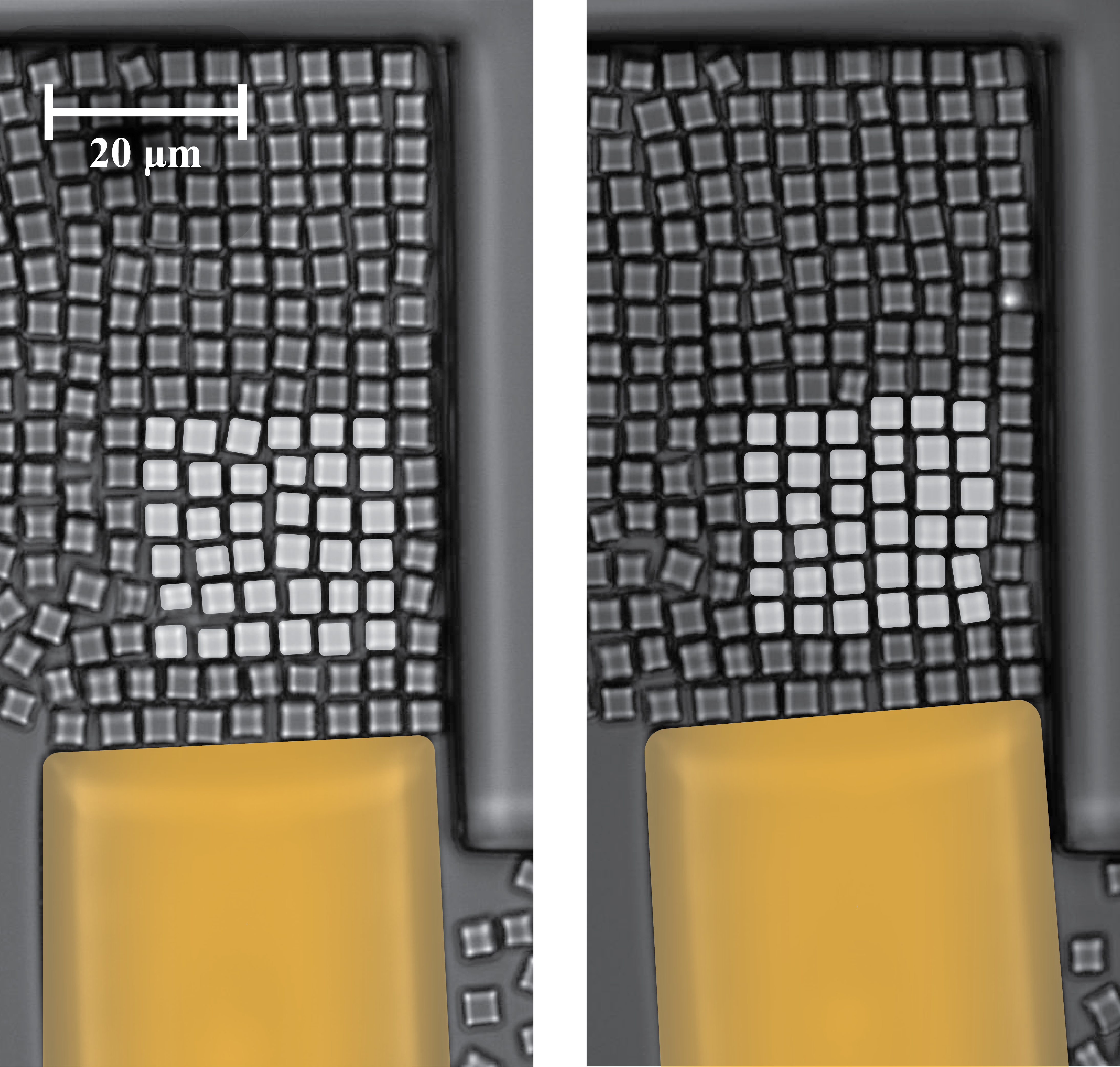}
\end{minipage}\hfill
\begin{minipage}[t]{0.25\linewidth}
\raggedright d)\\
\includegraphics[height=4.5cm]{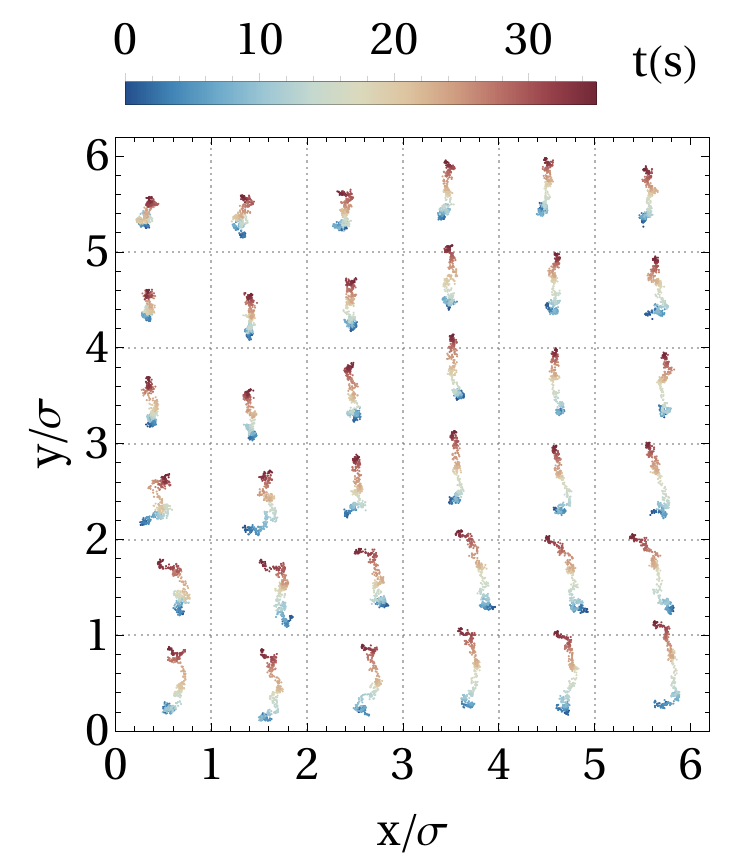}
\end{minipage}\hfill
\begin{minipage}[t]{0.45\linewidth}
\raggedright e)\\
\includegraphics[height=4.5cm]{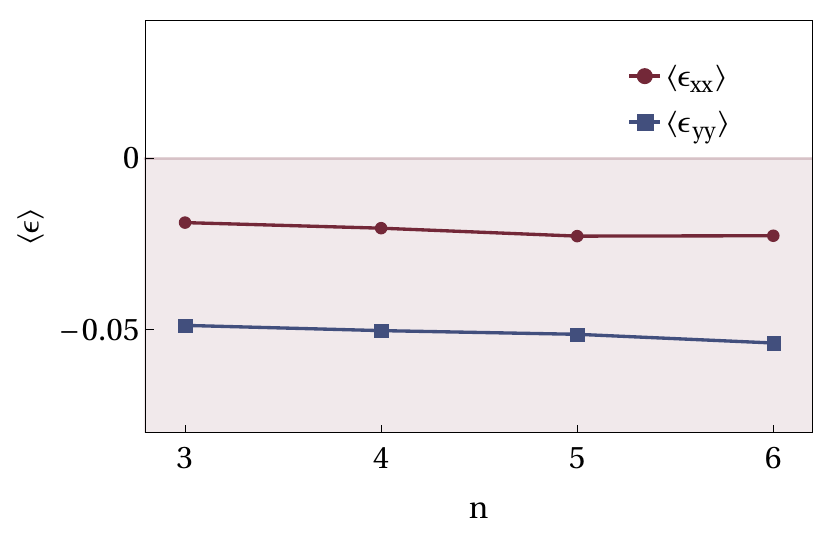}
\end{minipage}
\end{minipage}

\end{tcolorbox}
\vspace{0.3cm}

\begin{tcolorbox}[enhanced,
  colframe=black, colback=white,
  boxrule=0.4mm, arc=4mm,
  width=\linewidth, boxsep=2mm]

\begin{minipage}[c]{0.5cm}
\vfill
\rotatebox{90}{\centering III. GRANULAR EXPERIMENTS}
\vfill
\end{minipage}%
\hfill
\begin{minipage}[c]{0.96\linewidth}
\begin{minipage}[t]{0.30\linewidth}
\raggedright f)\\
\includegraphics[height=4.4cm]{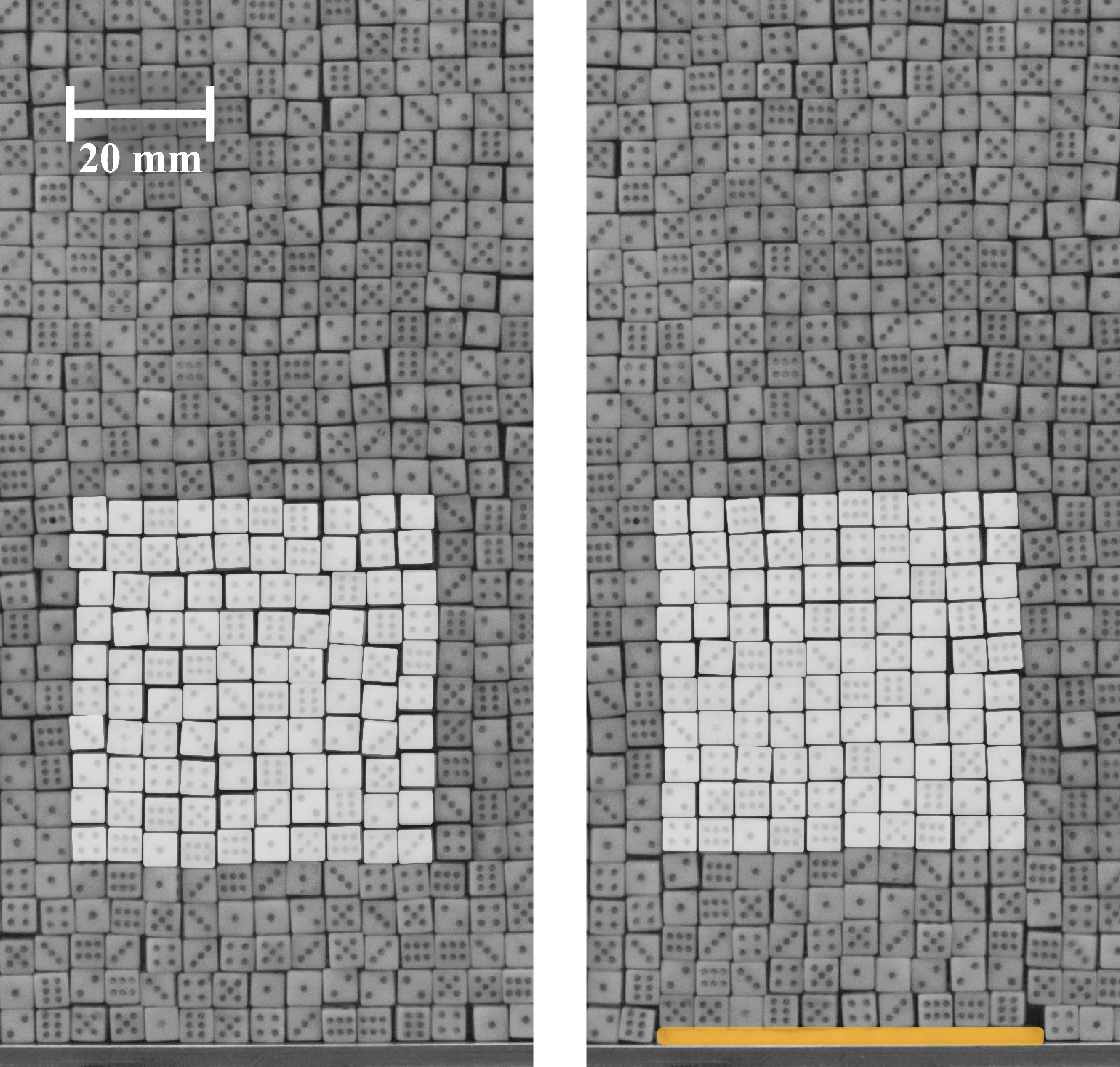}
\end{minipage}\hfill
\begin{minipage}[t]{0.25\linewidth}
\raggedright g)\\
\includegraphics[height=4.5cm]{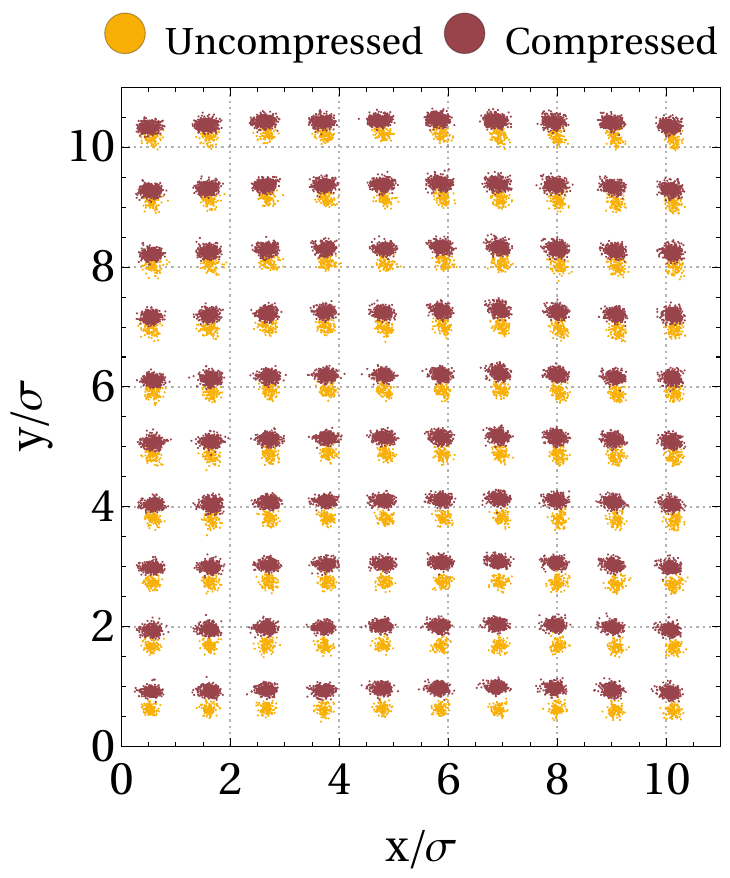}
\end{minipage}\hfill
\begin{minipage}[t]{0.45\linewidth}
\raggedright h)\\
\includegraphics[height=4.5cm]{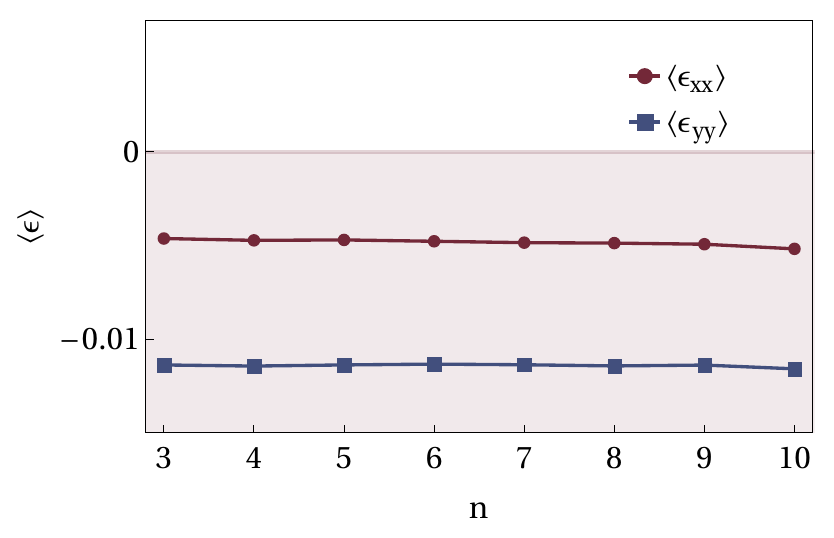}
\end{minipage}
\end{minipage}
\end{tcolorbox}

\caption{Three set-ups measuring the sign of the Poisson ratio for a (quasi) two-dimensional system consisting of squares. Set-ups consist of I. simulations, II. a colloidal experiment and III. a granular experiment. In all simulations/experiments, average coordinates are measured for a subset of the system (indicated by the white particles in panels a), c, and f)), both before and after applying an uniaxial compression. a),c) and f) show snapshots of the systems before and after the applied deformation. In all figures, the yellow objects are used to apply the uniaxial compression. d) shows the trajectories of the colloidal squares throughout the experiment and g) shows the coordinates before and after compression of the granular squares. Strain tensors are fitted for all contiguous $n\times n$ regions ($n\in[3, N]$) within the white subsystem. In panels b), e) and h)  $\langle \epsilon_{xx}\rangle(n)$ and $\langle \epsilon_{yy}\rangle(n)$ are plotted as a function of $n$ for the three systems respectively. The fact that in all three cases both strain-tensor elements are negative indicates a negative Poisson ratio.}
\label{fig:experiments}

\end{figure*}
\section{Experimental realization}
To demonstrate that the observed negative Poisson ratio is not just a feature of an artificial model system, we now explore the same Poisson ratio in two experimental systems: one on the colloidal and one on the granular scale. In both cases, we make use of a monolayer of hard cube-shaped particles placed on a flat substrate, such that the particles effectively act as hard squares. Since we cannot directly measure stress tensors experimentally in these systems, we first have to design a method to estimate the sign of the Poisson ratio.

The strategy that we use to determine the sign of the Poisson ratio is illustrated in Fig.~\ref{fig:experiments}a). Starting from an undeformed crystal of squares, we push locally on one row of particles (yellow in Fig.~\ref{fig:experiments}a)) and examine the behaviour of the white highlighted particles before and after this compression. If the Poisson ratio is negative, the horizontal spacing between the white squares should shrink as a result of the vertical compression. 

\begin{figure}[t]
    \centering
    \includegraphics[width=\linewidth]{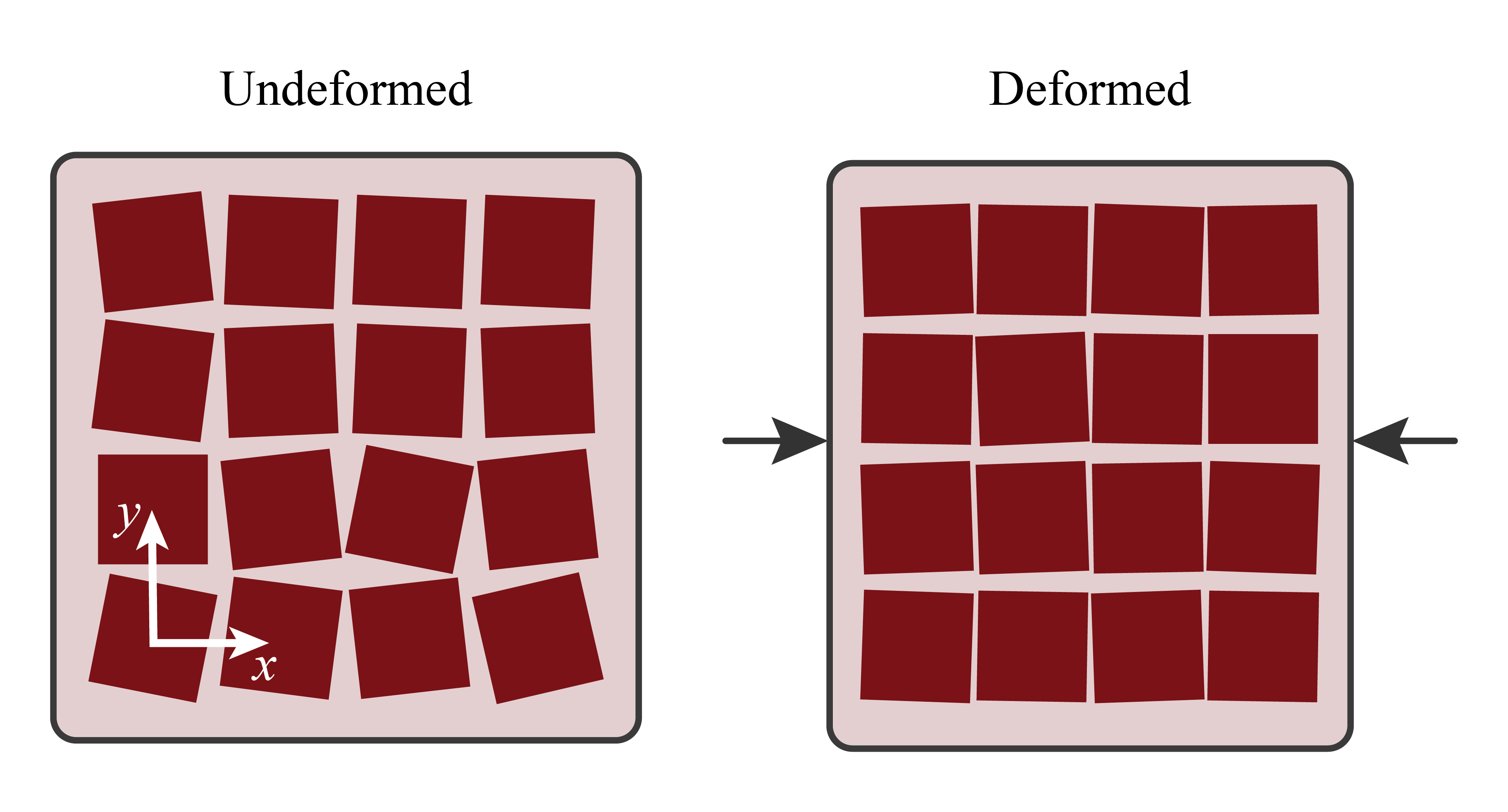}
    \caption{Cartoon of the mechanism behind the negative Poisson ratio. When the system is compressed along the $x$-axis, configurations in which particle faces align are entropically more favourable leading to stronger alignment. As a result, particles gain more translational freedom in the perpendicular direction reducing the transverse pressure.}
    \label{fig:mechanism}
\end{figure}

To quantify this effect, we measure the time-averaged coordinates before and after compression, $\{\langle\mathbf{r}^{uc}\rangle\}$ and $\{\langle\mathbf{r}^{c}\rangle\}$ respectively, for the $N\times N$ subsystem of white squares. Subsequently, we fit the compressed positions as a function of the uncompressed coordinates with a function of the form
\begin{align}
    \langle \mathbf{r}_{i}^{c}\rangle = \left(
\begin{array}{cc}
 d_{xx} & 0 \\
 0 & d_{yy}
\end{array}
\right) \cdot \langle \mathbf{r}_i^{uc}\rangle + \mathbf{c},
    \label{eq:deformationfit}
\end{align}
where $d_{xx}$ and $d_{yy}$ define a (diagonal) deformation matrix, and $\mathbf{c}$ corresponds to a center-of-mass shift of the subsystem.
Note that we neglect off-diagonal terms, i.e. we assume that compression does not lead to rotation. 
In order to check for effects of the size of the analyzed region, as well as any effects of nearby walls, we repeat this fit for 
 all contiguous $n\times n$ regions ($n\geq 3$) within the $N\times N$ subsystem. This yields $\langle d_{xx}\rangle(n)$ and $\langle d_{yy}\rangle(n)$, which we then use to determine the corresponding Lagrangian strains $\langle \epsilon_{xx}\rangle(n)$ and $\langle \epsilon_{yy}\rangle(n)$. Finally, we estimate the sign of the Poisson ratio from the ratio $-\epsilon_{xx}/\epsilon_{yy}$. Since $\epsilon_{yy}$ is always negative (due to the compression), we expect a negative Poisson ratio to correspond to the case where $\epsilon_{xx}$ is also negative. 

To validate this method, we first apply it to a hard-square system modeled via Monte Carlo simulations, in which a row of particles is pushed inward using an artificial external potential. Simulation details are explained in the Methods. 
In Fig.~\ref{fig:experiments}b), the strains $\langle\epsilon_{xx}\rangle$ and $\langle\epsilon_{yy}\rangle$ are plotted as a function of $n$. Since both strains are negative, this method predicts a negative Poisson ratio for the hard squares, providing strong evidence that the method works. 

We next apply this method to a colloidal system of 3D-printed hard cubes with an edge length of 4$\mu$m, where the crystal forms due to an added depletant, depicted in Fig.~\ref{fig:experiments}c) (see Methods for more system details). A large ``pusher''  (highlighted in yellow) is printed, and moved using optical tweezers to compress the crystal against a set of immobile walls that are printed on the substrate. As our subsystem, we choose a 6x6 region of particles in front of the pusher, avoiding defects and the wall (highlighted white particles). The trajectories of these particles during the push are shown in Fig.~\ref{fig:experiments}d). As shown in Fig.~\ref{fig:experiments}e) we find a negative Poisson ratio for this system.

To demonstrate the robustness of the phenomena to both out of equilibrium effects and changes in length scale, we perform analogous experiments on a granular system of hard cubes (6-sided dice with edge length approx. $0.5$mm) on a vibrating plate, as shown in Fig.~\ref{fig:experiments}. To deform the square crystal, we use a cardboard pusher (highlighted in yellow). As our subsystem, we choose a 10x10 region of dice, whose coordinates before and after compression we plot in Fig.~\ref{fig:experiments}g). We again find a negative Poisson ratio, as shown in Fig.~\ref{fig:experiments}h).

\begin{table}[!t]  
\centering
\begin{center}
\begin{tabular}{l @{\hskip 1cm} l }
\textbf{Crystal}&\textbf{Poisson ratios}\\
\hline\\
\begin{tabular}{l}Squares \\Triangles\\Hexagons\end{tabular} & $\nu_{(12)(21)} = \frac{B_{12}}{B_{11}}$\\
\\
\hline
\\
\begin{tabular}{l}Cubes\end{tabular} &$\nu_{(12)(21)(13)(31)(23)(32)} = \frac{B_{12}}{B_{11} +B_{12}}$\\
\\
\hline
\\
\begin{tabular}{l}Triangular prisms\\
Hexagonal prisms\end{tabular} &\begin{tabular}{l}$\nu_{(12)(21)}=\frac{B^2_{23} -B_{12}B_{33}}{B^2_{23} -B_{11}B_{33}}$\\$\nu_{(13)(23)} =\frac{(B_{11} -B_{12})B_{23}}{-B^2_{23} +B_{11}B_{33}}$\\$\nu_{(31)(32)} =\frac{B_{23}}{B_{11} +B_{12}}$\end{tabular}\\
\\
\hline
\end{tabular}
\caption{Independent Poisson ratios, expressed in terms of the elastic constants, for the various hard-polyhedra crystals.}
\label{tabel:poissonratiocrystal}
\end{center}
\end{table}

\section{An intuitive explanation}

Since we have established a negative Poisson ratio for both simulations and experiments on hard squares, a natural follow-up question is: what mechanism gives rise to this behaviour? In the systems we study, interactions are based purely on excluded volume, meaning that for the systems that are in thermal equilibrium, all behaviour emerges from the system maximizing its entropy under the geometric constraints imposed by particle shape. When the system is compressed along an axis, configurations in which particle faces align are entropically more favorable leading to stronger alignment. For the crystal of squares, it is easy to see that, as particles align along the compression axis, they gain more translational freedom in the perpendicular direction (see Fig.~\ref{fig:mechanism}). This reduces the transverse pressure and results in a negative Poisson ratio. This simple entropic argument can be made more quantitative using cell theory, where the free energy of the crystal is approximated based on the motion of a single particle in a static cage of its neighbours.

\begin{widetext}

\begin{figure*}[t]
        \begin{minipage}[t]{\linewidth}
        \vspace{0pt}
        \includegraphics[width=\linewidth]{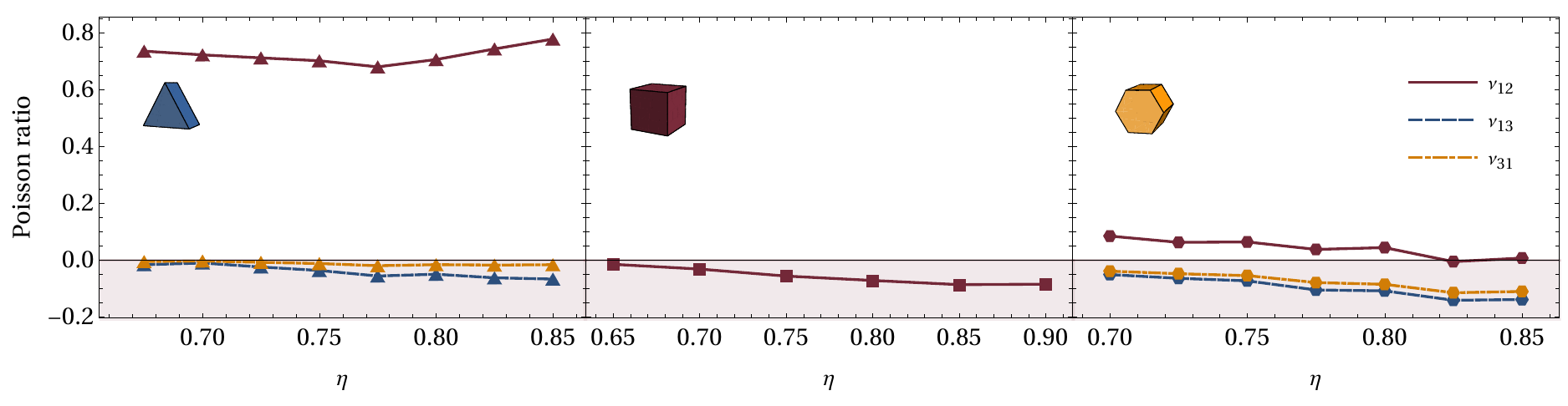}
        \end{minipage}
    \caption{Poisson ratios for the three three-dimensional systems considered in this work, plotted as a function of packing fractions. Note that for cubes, $\nu_{12} = \nu_{13} = \nu_{31}$. Full measurements are provided in the Supplementary Information.}
    \label{fig:3Dpoisson}
\end{figure*}
\end{widetext}
\par
\noindent
Indeed, as we show in the SI, this calculation yields a negative Poisson ratio for hard squares. Although the granular systems are not in equilibrium, we expect similar arguments to hold. 

\section{Elastic behaviour of crystals of hard polygons and polyhedra}

As the behaviour we observe in the square system is associated with the alignment of flat planes, it is intriguing to wonder whether a negative Poisson ratio is a generic feature of hard polyhedra. To this end, we calculate Poisson ratios for a range of 2D polygons and 3D polyhedra that can fill space, see Figs. \ref{fig:difcrystals}a)-f). We focus on the Poisson ratios associated with the cardinal directions ($x$, $y$, $z$) as depicted in Fig.~\ref{fig:difcrystals}.  Note that depending on the shape, the number of independent elastic constants and Poisson ratios changes. An overview is shown in Table \ref{tabel:poissonratiocrystal}. Full details how the effective elastic constants are measured in these systems can be found in the SI.

The 2D results for the Poisson ratio $\nu_{12}$ corresponding to the directions depicted in Fig.~\ref{fig:difcrystals} are shown in Fig.~\ref{fig:2Dresultstog}b). These 2D systems appear to span a wide range of behaviours, with triangles having a large positive Poisson ratio and hexagons being positive but close to 0. The large positive nature of the triangle system likely arises due to triangles acting as wedges, pushing each other apart in the perpendicular direction while being compressed. Hence, face alignment alone is not a sufficient condition for auxetic behaviour. The 3D systems, shown in Fig.~\ref{fig:3Dpoisson} display similar behaviour, clearly exhibiting a competition between flat face alignment favoring a negative Poisson ratio, and wedge-induced pushing towards a positive Poisson ratio. 

We stress that for all systems we investigate here, we observe at most \textit{partial} auxeticity: the sign of the Poisson ratio depends on the direction in which the compression is applied. This is already clear for the triangular and hexagonal prisms from the data in Fig.~\ref{fig:3Dpoisson}, but also holds for the cubes and squares, which have a positive Poisson ratio if a compression is applied along a direction that deviates sufficiently from the cardinal lattice vectors. As an example, in the SI we show explicitly how the Poison ratio for squares and cubes changes as we rotate the system around the z-axis, resulting in values significantly above zero.

\section{Conclusion}
In conclusion, we have found negative Poisson ratios in various entropy-driven systems of polyhedra. Predicting which shapes will exhibit (partial) auxeticity is in general non-trivial, yet our work demonstrates that auxetic behaviour can emerge in very simple systems from an interplay between geometry and entropy alone. Moreover, we have demonstrated that the phenomenon is experimentally realizable and robust, by showing that it appears both in systems of colloidal particles driven by Brownian motion and granular particles driven by external vibrations.

\section*{Data Availability Statement}
Data associated with the measurements, as well as all figure data, will be published online prior to publication of the paper.

\section*{References}
\addcontentsline{toc}{part}{Bibliography}
\markboth{\MakeUppercase{Bibliography}}{}
\bibliographystyle{abbrvunsrt2}
\bibliography{refPR}

\section{Methods}

\subsection{Simulation details for measuring elastic constants}

In order to measure the stress tensors required to obtain the Poisson ratios, we performed simulations of both the equilibrium crystals and the deformed crystals under consideration. The crystals were simulation using event-driven molecular dynamics (EDMD) simulations \cite{Rapaport2009,smallenburg2022efficient}  in the canonical ensemble (i.e. at fixed number of particles $N$, volume $V$ and temperature $T$) with periodic boundary conditions. All simulations were started from defect-free crystals. Each crystal contained approximately $1000$ particles, where the number of crystal layers in each direction was chosen such that the undeformed box was as square/cubic as possible. For the hexagonal- and triangular prism crystal, the lattice constants were chosen such that the undeformed system exhibited isotropic pressure (see SI). Each simulation ran for $t/\tau = 5\times 10^5$ (2D) or $t/\tau = 2.5\times 10^5$ (3D), with $\tau = \sqrt{\frac{m\sigma^{d}}{k_BT}}$, with $k_B$ is Boltzmann's constant, $m$ is the particle mass, $\sigma$ the relevant lengthscale and $d$ the dimension. For each crystal, we considered a range of different packing fractions $\eta$, in which the crystals did not melt, rotate under deformation.
The stress-strain curves were fitted up to $\mathcal{O}(\delta^2)$, where we included values of $\delta\in[0, 0.008]$ for the square crystal and  $\delta\in[0, 0.01]$ for all other crystals. For the square crystal, larger values of $\delta$ occasionally caused the crystals to rotate.

\subsection{Details simulation mimicking experiments}
We used a Monte Carlo (MC) simulation in the $NVT$-ensemble to test the strategy to determine the sign of the Poisson ratio in the colloidal and granular experiments. The system consisted of  
$N=900$ squares in a $30\times30$ square at $\eta\approx0.9$ and the simulations was implemented with hard-wall boundary conditions. After measuring the average coordinates in the undeformed system, the middle $N_\text{push} = 9$ particles at one side of the system were pushed inwards (see Fig.~\ref{fig:experiments}a),b)) to deform part of the system. To achieve this, we applied a harmonic trap to these nine particles of $ V = \frac{1}{2} k\left(\frac{y-y_0}{r^\text{outer}}\right)^2$, with reduced stiffness $ k = 100k_BT$ and center position $y_0 = (\frac{1}{\sqrt{2}} + 1.25)r^\text{outer}$, where $r^\text{outer}$ denotes the outer radius of the squares. 

Both before and after pushing, the system equilibrated for $5\cdot 10^5$ cycles. Average (un)deformed coordinates were measured during $2\cdot 10^4$ cycles, were measurements were performed every $\Delta n= 100$. 

\subsection{Colloidal experiments}

\subsubsection{Materials}
Poly(ethelene glycol) (PEG, M\textsubscript{v} 600,000), Novec 7100 Engineering Fluid (HFE-7100) and propylene glycol monomethylether
acrylate (PGMEA, $99.5\%$) were purchased from Sigma-Aldrich. 2-propanol (IPA,9, $99\%$)
was purchased from VWR Chemicals. Poly(vinyl alcohol) (PVA, MW 78000, $88\%$ hydrolized) was purchased from Polysciences. All materials were used as received unless stated otherwise. All aqueous solutions were prepared from deionized water with 18.2M$\Omega$ cm resistivity, using a Millipore Filtration System (Milli-Q Gradient A10).

\subsubsection{Design}
The experimental set-up needed to consist of the following key components: (i) a non-mobile confining structure that provided boundaries to our system, (ii) a sliding block to generate the deformation on the crystal using manipulation by optical tweezers, and (iii) the cubes. The desired microscopic system thus required both mobile and non-mobile 3D printed features, as can be seen in Fig.~ \ref{fig:fig_silvana_1} A. We realized these three elements by executing three consecutive steps: first, the immobile confining structures were printed on a glass slide. After development of the structures, the slide including the confining structure was spin-coated with a thin layer of PVA, followed by a second printing step. Any structures printed on top of the PVA layer could subsequently be detached by dissolution of the PVA upon addition of water, while all structures printed on the glass slide in the first step remained attached. The rationale for the design of the three different elements is described below.\\

\textbf{Non-mobile confining structure.} We designed a set of walls that functions as rigid boundaries during the measurement. These structures are represented in blue in Fig.\ref{fig:fig_silvana_1} A. The walls have right-angled corners to facilitate the confinement of mobile structures after their release from the substrate. To ensure that this structure remained intact after spin-coating the PVA layer, its contact area with the substrate was maximized, and the wall height was kept at a minimum of 10 $\mu m$. Additionally, a set of openings was introduced to drain any excess polymer to avoid inhomogeneities in the polymer concentration with ensuing osmotic pressure differences and fluid flows, as well as prevent the formation of bubbles. The inner dimensions were chosen to fit as many mobile features as possible, and to facilitate interface finding during the second printing step by allowing sufficient distance between the walls and the mobile structures.\\

\textbf{Sliding blocks.} The purpose of these blocks was two folded: (i) to push cubes together and towards the corners to promote crystal formation close to the confinement boundaries and (ii) as a mean to perform small unilateral deformations on the crystal. For the latter, it was important that the blocks' shorter axis was in the $xy$-plane to avoid unwanted alignment with the optical trap. Three different types of blocks were included in the design, each with different dimensions and shapes (see Fig.~\ref{fig:fig_silvana_1} A), to provide alternatives for moving and pushing the cubes.\\

\textbf{Cubes.} The cubes were designed with a side length of 4 $\mu m$, corresponding to the minimum printable dimension that ensures sharp edges and flat faces. This size also allowed for pronounced Brownian motion and increased number of particles inside the confining structure. Arrays of cubes were distributed at equal distances close to the non-mobile confinement to improve the chances of crystal formation after release from the substrate.

\subsubsection{3D printing of micro-structres}
\begin{figure*}[p]
    \centering
    \includegraphics[width=\linewidth]{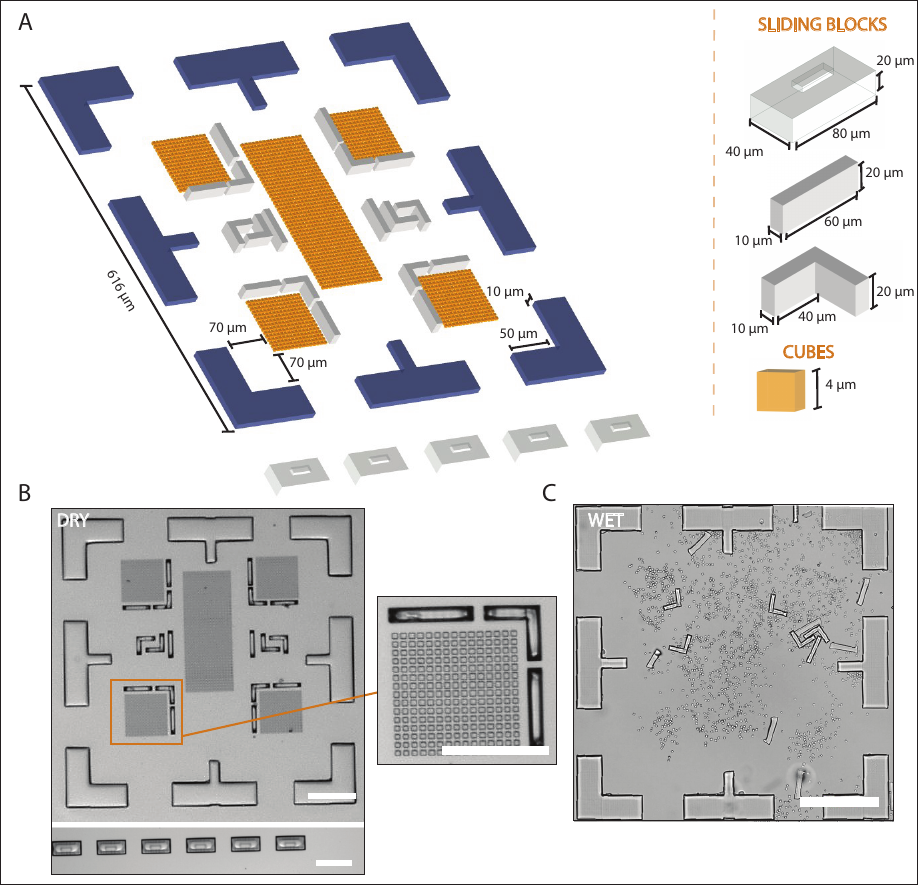}
    \caption{\textbf{Experimental design of the colloidal experiment.} \textbf{(a)} AutoDesk Inventor render of the assembled design consisting of the non-mobile confining walls (blue), the sliding blocks (gray) and the cubes (orange), accompanied by relevant dimensions of the features. \textbf{(b)} Brightfield microscopy images of the print before addition of the depletant solution. \textbf{(c)} Brightfield microscopy images of the print after addition of the depletant solution. Note that the confining walls stays on the surface while the cubes and the sliding blocks become detached. Scale bars are 100 $\mu m$.}
    \label{fig:fig_silvana_1}
\end{figure*}

3D micro-printing was employed for both the realization of the non-mobile walls, as well as for the mobile cubes and the sliding blocks. All the structures were designed in Autodesk Inventor, and rendered in Describe. We used a commercially available two-photon polymerization (2PP) 3D micro-printer (Photonic Professional Gt, Nanoscribe GmbH).\\
\textbf{Non-mobile structure}
Using a 25x oil-immersion objective (Zeiss, NA=0.8) the non-mobile confining structure was printed on top of a cover-slip (25 mm No.1) with IPS photoresist (Nanoscribe GmbH). After printing, the structures were developed by a 30 min submersion in PGMEA, and a 2 min dip into IPA.
\textbf{Mobile cubes and sliding blocks}
A $7\%w/w$ solution of PVA (M\textsubscript{v} 600,000) was spin coated on top of the cover-slip with the already existing non-mobile confining structure, at 4000 rpm for 60 s. The cover-slip was then baked on a hot plate at $80^\circ\mathrm{C}$ to dry the polymer layer. In preparation for printing, the cover-slip was aligned such that the angle between the $y$-axis of the walls and the x-axis of the holder was $\theta=90^\circ$. A thin silicon sheet with a hole in the middle was placed around the printing surface to trap the photo-resist and improve the interface finding, which is otherwise challenging in the presence of the PVA layer. Using a 63x oil-immersion objective  (Zeiss, NA=1.4) the mobile features were printed with IPL photo-resist. After printing, the structures were developed by a 30 min submersion in PGMEA, and a 2 min dip into HFE-7100 to avoid pattern deformation by capillary forces that occur during drying. Fig.~\ref{fig:fig_silvana_1} B displays a brightfield image of the final print.

\clearpage

\subsubsection{Depletion interactions for crystal formation}
For experimental observation, the coverslip with the confining structure, cubes, and sliding blocks was placed in a custom made microscope holder. 3 mL of an aqueous solution of 0.027-0.030 g/L PEG  was prepared by dilution from a 0.1 g/L stock. The aqueous PEG solution serves to detach the mobile features by dissolving the PVA on the surface, and acts as a depletant to promote crystal formation. Experimentally, the addition of a non-adsorbing polymer to induce depletion interactions among hard cubes has been documented in literature \cite{Rossi2011, Rossi2015}. The effective attraction arises from a polymer exclusion layer around each cube, where the layer has a characteristic thickness set by the radius of gyration $R_g$ of the polymer. When two exclusion zones overlap, the volume available to the polymers increases, which increases the entropy of the system. The attractive interaction potential between two cubes can be calculated as\cite{Asakura1954,Lekkerkerker2024,Vrij2009} : $U\approx-\Delta V n k_BT$ where $n$ is the number density of depletants and $\Delta V$ is the change in volume excluded to the polymers because of overlapping exclusion zones, i.e. for two perfectly aligned and touching cubes  $\Delta V\approx(L +2R_g)^22R_g$ where $L$ is the side length of the cube\cite{Rossi2011}. For our system this yields a contact value of the attractive potential of  $U_\mathrm{contact}\approx -28k_BT$ considering $R_g=28.68$ nm based on the measurements for PEG 600K by Ziebackz \textit{et  al.}\cite{Ziebacz2011} . 
Once the PEG solution is added into the microscope holder, the cubes and sliding blocks immediately detach from the surface while the confining structure stays (see Fig.~ \ref{fig:fig_silvana_1} C). After one minute, all mobile structures have settled again onto the substrate due to sedimentation. It is important that the addition of the PEG happens in 100 $\mu L$ steps to minimize disturbance of the structures.

\subsubsection{Optical manipulation and observation}
 Particles were imaged using a Nikon Eclipse Ti microscope with 60x water immersion (NA=0.7) objective. An optical trap was generated by an in-house setup that uses a highly focused laser beam (1064 nm Nd:YAG laser, LaserQuantum) integrated into the light path to simultaneously image and trap. The system was allowed to evolve for approx. 10 minutes until enough particles collected in the same region. As the cubes did not necessarily sediment close to one of the non-mobile confining corners, it was necessary to push them to such an area using the sliding blocks. Additional time was allowed for the structure to crystallize after the disturbance, and defects were manually fixed by removing or placing individual cubes into/out of the structure. 
 
 Once the desired crystal size and shape was achieved, the recording was started. For the first 15 s of the recording, no pressure was exerted on the crystal by the sliding blocks. After 15 s, we used the optical trap to push the shorter edge of a sliding block (40 $\mu m$ x 80 $\mu m$) against the crystal, parallel to the confining side to create the necessary deformation (see Fig.~
 \ref{fig:experiments}d)). The block was pushed slowly and only for a few microns to not disturb the crystalinity too much. After the compression, some more frames were recorded in order to get the information on the cubes new positions in equilibrium. The experiment was recorded at 20 FPS.

 \subsubsection{Data Analysis}
The centers of mass of all cubes inside the crystal were tracked using a custom Python routine that pre-processes the image by using a binary filter and removing thin bridges and artifacts between features, followed by edge and area detection employing a Gaussian filter and size restrictions. Coordinates were linked over time into particle trajectories using the Trackpy\cite{trackpy2023} algorithm, and a static feature visible in the field of view was used to correct for drift. The drift-corrected trajectories are used for the analysis.

\begin{figure}[t]
        
      \begin{minipage}[t]{\linewidth}
        \vspace{0pt}
        \includegraphics[width=\linewidth]{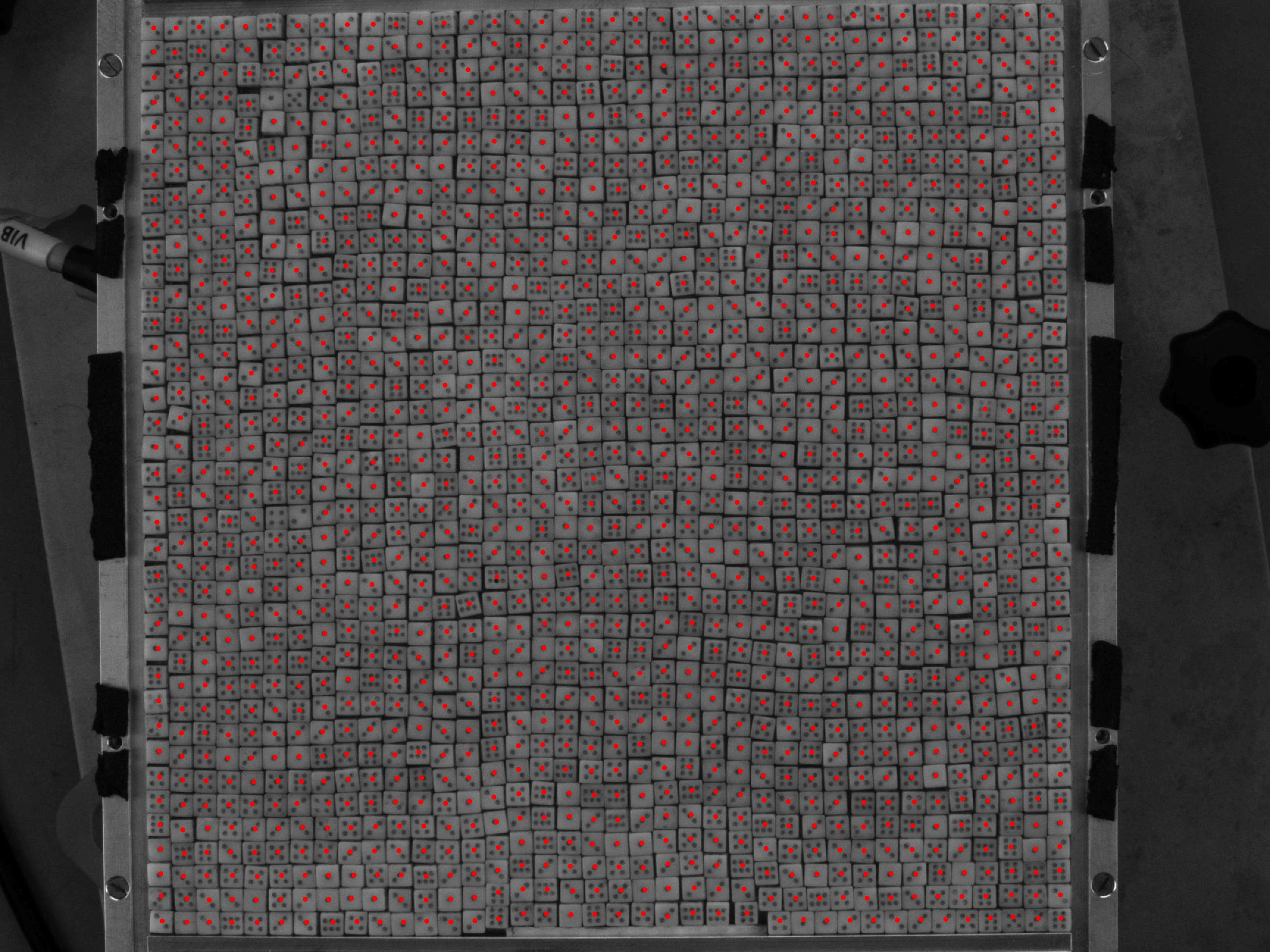}
        \end{minipage}
    
    \caption[width=1\linewidth]{Granular system of dice, with a cardboard pusher (bottom side) compressing the middle ten rows. The red dots indicate the tracked particle coordinates. }
    \label{fig:granulardice} 
\end{figure}
\subsection{Granular experiments}

Our granular setup consists of granular cubes on a vibrating plate. Specifically, we use the same setup as in Ref. \onlinecite{plati2024quasi}, but instead of spherical particles, we use small acrylic dice with an edge length of 5 mm. These dice are placed on a squared aluminium plate (1 cm thick) which is vertically vibrated by an electrodynamic shaker (Brüel \& Kjær, LDS V400). The particles are confined in a square region (side length $L=20cm$) by aluminium bars attached to the bottom plate. The particles are not confined from above except by gravity. The system is imaged from above using a high-resolution (5 MP) camera (Basler a2A2590), using a lens lens (Basler C125-1218-5M-P) with a fixed focal length of 12.0 mm. Images of the system are taken at a frame rate of two frames per second. The setup is illuminated by four light panels  placed around the apparatus. 

To initialize the system, particles were placed on the plate in a 38x38 square lattice in a single layer, corresponding to a packing fraction of approximately 0.90. We then turn on the shaker to impose a sinusoidal vibration at a frequency of 210~Hz and a maximum acceleration of $8.5g$, with $g$ the gravitational acceleration. This provides enough energy for the particles to move and collide within the quasi-2D confinement without bouncing up high enough to escape the monolayer. We let the system shake for several minutes to obtain baseline data for an unperturbed crystal. To locally compress the system along one of the lattice directions, we insert a flat piece of a cardboard between the bottom plate and the confining bar on one side of the system, pushing the particles on that side of the box inwards (see Fig.~\ref{fig:granulardice}). The cardboard pusher is then kept in a fixed position for several minutes to gather data for the compressed case. 

From the captured images, approximate particle coordinates are obtained using basic image analysis functions in Wolfram Mathematica. A typical example of tracked coordinates is shown in Fig.~\ref{fig:granulardice}. Note that while occasionally the coordinate detection misses particles when they are very close together or close to the edge of the system, in practice the vast majority of particles on the interior of the sample are accurately located.

\end{document}